\documentstyle[PASJadd]{PASJ95}

\begin{document}

\title{Reply to 
``Roles of SNIa
 and SNII in ICM Enrichment'' by\\ Y. Ishimaru and N. Arimoto}
\author{Michael {\sc Loewenstein}\thanks{%
also with the University of Maryland Department of Astronomy}
and Richard F. {\sc Mushotzky}\\
{\it Laboratory for
High Energy Astrophysics, NASA/GSFC, Code 662, Greenbelt, MD 20771}\\
{\it E-mail(ML): loewenstein@lheavx.gsfc.nasa.gov}}

\abst{We address a number of misunderstandings and misstatements
contained in the paper ``Roles of SNIa and SNII in ICM Enrichment'' 
by Y. Ishimaru and N. Arimoto with regard to the papers by
Mushotzky et al. (1996, ApJ, 466, 686) and 
Loewenstein and Mushotzky (1996, ApJ, 466, 695). In particular, we emphasize
that comparison between observations and models
in these papers were made self-consistently
assuming a particular ({\it i.e.}, photospheric) 
value for the solar iron abundance. We also briefly revisit the
question of the contribution of Type Ia supernovae to the iron
enrichment of the intracluster medium.}

\kword{Clusters of galaxies: intracluster medium -- Clusters of 
galaxies: X-rays -- Galaxies: intergalactic medium -- Galaxies: abundances}

\maketitle

\section{Introduction}

The origin of heavy elements in the intracluster medium (ICM) is an 
important area of astrophysical research, with implications for
the understanding of galaxy and cluster formation,
structure, and evolution. We were principle authors on
a pair of papers (Mushotzky et al. 1996, Loewenstein and Mushotzky 1996)
addressing this issue using {\it ASCA} data of four clusters of galaxies.

The first paper (Mushotzky et al. 1996) presented the results of
the analysis of {\it ASCA} spectra, noted the superficial
resemblance of the abundance pattern to that calculated for
Type II supernovae (SNII),
and briefly commented on the
large number of SNII required to account
for these observations
and the resulting implications for our understanding of
primordial star formation in proto-elliptical galaxies.

The second paper (Loewenstein and Mushotzky 1996) presented a more detailed
comparison of the observed elemental mass-to-light ratios
and abundance ratios
with models of enrichment from SNII explosions
of stars with masses distributed according to 
power-law initial mass functions (IMFs) of various slope. Mass-to-light ratios 
for Fe, Si, O, etc. were derived by integrating nucleosynthetic
yields calculated by Woosley \& Weaver (1995) over the IMF, assuming that
the present-day light is emitted by stars from the same IMF
that produced the SNII, and ignoring Type Ia supernovae (SNIa). We found that
a flat IMF was required to reproduce the observed mass-to-light ratios,
and that the results were consistent for Fe, Si, and O. Since O is
exclusively, and Si predominantly, synthesized by SNII there is no
need to invoke significant enrichment by SNIa. It, of course,
does {\it not} follow that the importance of SNIa in accounting for
the Fe enrichment can be ruled out, and we did not do so (see below).

We also compared observed ratios of abundances relative to Fe
with abundance ratios in the models. This comparison was made in
an internally consistent way
by forming the ratios of the model mass-to-light ratios and then
{\it normalizing them to the same abundances taken to be solar in the spectral
analysis}. Because of uncertainties in the measured abundances
and in the nucleosynthetic Fe yields, we did not draw strong conclusions
about the IMF from this comparison.

An additional unknown that plays off against the Fe yield
uncertainty in explaining
abundance ratios relative to Fe, 
is the contribution of SNIa to the Fe enrichment.
As stated in Section 7.2 of Loewenstein \& Mushotzky (1996), 
more than one-third of the Fe could originate from SNIa if the
standard Woosley \& Weaver (1995) Fe yield is assumed; moreover, this
fraction can climb to 75\% or more if the yield uncertainty is
explicitly considered. The ambiguity in interpreting the observed
Fe abundances has little effect on our conclusions about the number of
SNII required to explain the observations since large O, Ne, and
Si abundances are also observed.

\section{Discussion}

In the paper by Ishimaru \& Arimoto (1997) it is implied
that we were inconsistent in using photospheric abundances
in data analysis while referring to models that assumed
meteoritic abundances, and that we ruled out the possibility of
significant 
ICM Fe enrichment by SNIa. As should be clear from the above discussion,
neither of these is the case.

First and foremost, our primary conclusions are based on a comparison of
observed elemental mass-to-light ratios with those predicted by the models --
the assumed solar abundances nowhere enter into calculation of the latter.
Secondly, as explained above, we normalized all abundance ratios
to the photospheric values -- even if the
meteoritic abundances are correct, as long as consistency is maintained
the comparison is valid. We found that a bimodal star formation
model with an IMF slope of $\sim 1.5$ could explain the observed ratios, but
noted that this depended on the uncertain assumptions that SNIa were
negligible and that nucleosynthetic yields of Fe are precisely known.
Timmes, Woosley, \& Weaver (1995) have noted that the latter may be
uncertain by a factor of two.

As repeatedly stated in Loewenstein \& Mushotzky (1996), the 
observed abundance ratios neither require nor rule out an
important SNIa contribution to the Fe
enrichment of the ICM. We have revisited this question
in more detail for the clusters AWM 7 and Abell 1060, deriving 
best simultaneous fits of the total numbers of SNIa and SNII
to the observed abundance pattern of Fe, O, and Si. 
Yields for SNIa are taken from Model W7 of Nomoto et al. (1984), and 
SNII yields either from
Woosley \& Weaver (1995; model WW-1
in Loewenstein \& Mushotzky 1996), or Thielemann, Nomoto, \& Hashimoto (1996).
SNII yields are averaged 
over a Salpeter IMF with an upper limit of 40M$_{\odot}$.
We note that the Woosley \& Weaver (1995) yields include unprocessed ejecta
as well as newly synthesized SNII products.
The best-fit values of the mass fractions in the ICM from SNIa are
displayed in Tables 1 and 2 for Fe, Si, O, and for the
sum of all observed elements (O, Ne, Mg, Si, S, Ar, Ca, 
and Fe).
Upper and lower limits are crudely and
conservatively estimated by allowing the abundances
to independently vary within their 90\% confidence limits.
It is clear that there is a large dependence on which set of
SNII yields are adopted: the Woosley \& Weaver (1995) models favor a modest
SNIa contribution to the Fe and negligible contribution
to the Si enrichment while not ruling out pure SNII enrichment;
the Thielemann et al. (1996) yields (as discussed by Ishimaru \& Arimoto 1997)
favor a dominant contribution
to the Fe and
significant contribution to the Si enrichment from SNIa while
ruling out pure SNII enrichment.

\begin{table*}
\small
\begin{center}
Table~1.\hspace{4pt}Mass Fractions from SNIa$^*$.\\
\end{center}
\vspace{6pt}
\begin{tabular*}{\textwidth}{@{\hspace{\tabcolsep}
\extracolsep{\fill}}p{12pc}ccccc}
\hline\hline\\[-6pt]
Cluster & iron & silicon & oxygen & total\\
[4pt]\hline\\[-6pt]
Abell 1060 & 0.23($<0.74$) & 0.064($<0.40$) & 0.005($<0.045$) 
	   & 0.028($<0.22$)\\
AWM 7      & 0.22($<0.73$) & 0.062($<0.39$) & 0.005($<0.044$) 
	   & 0.027($<0.21$)\\
\hline
\end{tabular*}

\vspace{6pt}

\noindent
$^{*}$Woosley \& Weaver (1995) SNII yields are assumed.

\end{table*}

\begin{table*}
\small
\begin{center}
Table~2.\hspace{4pt}Mass Fractions from SNIa$^*$.\\
\end{center}
\vspace{6pt}
\begin{tabular*}{\textwidth}{@{\hspace{\tabcolsep}
\extracolsep{\fill}}p{12pc}ccccc}
\hline\hline\\[-6pt]
Cluster & iron & silicon & oxygen & total\\
[4pt]\hline\\[-6pt]
Abell 1060 & 0.68(0.39-0.92) & 0.32(0.12-0.71) & 0.017(0.005-0.082) 
           & 0.096(0.030-0.35) \\
AWM 7      & 0.64(0.27-0.90) & 0.20(0.048-0.56) & 0.014(0.003-0.072) 
           & 0.081(0.018-0.32) \\
\hline
\end{tabular*}

\vspace{6pt}

\noindent
$^{*}$Theilemann et al. (1996) SNII yields are assumed.

\end{table*}

\section{Conclusions}

The photospheric abundances adopted in
the spectral analysis of intracluster gas in four clusters
by Mushotzky et al. (1996)
were either irrelevant or self-consistently accounted for in their
conclusions and those in Loewenstein \& Mushotzky (1996).
Because of measurement error and uncertainties in the nucleosynthetic
yields of heavy elements by SNII (and the ironic fact that the best
measured element, iron, has the most poorly understood yield)
relative abundances of intracluster metals as measured by {\it ASCA}
do not, by themselves, 
significantly constrain the IMF of the SNII progenitors or the
relative contributions of SNIa and SNII to the enrichment of iron.
As pointed out by Loewenstein \& Mushotzky (1996)
and again in this work, and
emphasized by Ishimaru \& Arimoto (1997), SNIa may account
for 50\% or more of the Fe enrichment. However, any conclusions about
the SNIa contribution are highly dependent
on assumptions about nucleosynthetic yields (Gibson, Loewenstein, \&
Mushotzky, in preparation), and much smaller values are
allowed.
It is clear that 
virtually all of the observed O, and most of the observed Si do
originate from SNII, and that these number greatly in excess of what can
be produced by a simple stellar population with a standard IMF
(Loewenstein \& Mushotzky 1996).
\par
\vspace{1pc}\par
We thank B. Gibson for helpful discussions.

\section*{References}
\small
\re
Ishimaru, Y. \& Arimoto, N. 1997, PASJ, in press
\re
Loewenstein, M., \& Mushotzky, R. F. 1996, ApJ. 466, 695
\re
Mushotzky, R. F., Loewenstein, M., Arnaud, K. A., Tamura, T., Fukazawa, Y.,
Matsushita, K., \& Kikuchi, K. 1996, ApJ, 466, 686
\re
Nomoto, K., Thielemann, F.-K., \& Yokoi, K. 1984, ApJ, 286, 644
\re
Thielemann, F.-K., Nomoto, K.\& Hashimoto, M. 1996, ApJ, 460, 408
\re
Timmes, F. X., Woosley, S. E., \& Weaver, T. A. 1995, ApJS, 98, 617
\re
Woosley, S. E., \& Weaver, T. A. 1995, ApJS, 101, 181

\label{last}
\end{document}